\documentclass[
reprint,
superscriptaddress,
amsmath,amssymb,
aps,
pra,
floatfix,
]{revtex4-2}

\def\final{0}

\usepackage{braket}
\usepackage{mathtools}
\usepackage{graphicx}
\usepackage{color, xcolor}
\usepackage{dcolumn}
\newcolumntype{z}[1]{D{.}{.}{#1}}
\usepackage{physics}
\usepackage{subfigure}
\usepackage{tabularx}
\usepackage{hyperref}
\newcolumntype{Y}{>{\centering\arraybackslash}X}

\usepackage[linesnumbered,ruled ]{algorithm2e}
\SetKwInput{KwOutput}{Output}
\SetKwInOut{Input}{Input}
\SetKwInput{Output}{Output}
\UseRawInputEncoding 

\usepackage{xargs}
\usepackage[colorinlistoftodos,prependcaption]{todonotes}

\definecolor{Asparagus}{rgb}{0.53, 0.66, 0.42}
\definecolor{cornflowerblue}{rgb}{0.39, 0.58, 0.93}
\definecolor{darkolivegreen}{rgb}{0.33, 0.42, 0.18}
\definecolor{awesome}{rgb}{1.0, 0.13, 0.32}

\newtheorem{definitionenv}{Definition}
\newtheorem{lemmaenv}[definitionenv]{Lemma}
\newtheorem{theoremenv}[definitionenv]{Theorem}
\newtheorem{corollaryenv}[definitionenv]{Corollary}
\newtheorem{propositionenv}[definitionenv]{Proposition}
\newtheorem{conjectureenv}[definitionenv]{Conjecture}
\newtheorem{remarkenv}[definitionenv]{Remark}
\newenvironment{remark}{\begin{remarkenv}\rm}{\end{remarkenv}}
\newcommand{\br}{\begin{remark}}
	\newcommand{\er}{\end{remark}}

\newtheorem{exampleenv}{Example}
\newtheorem{app-lemmaenv}[section]{Lemma}

\newenvironment{definition}{\begin{definitionenv}\rm}{\end{definitionenv}}
\newenvironment{lemma}{\begin{lemmaenv}\rm}{\end{lemmaenv}}
\newenvironment{theorem}{\begin{theoremenv}\rm}{\end{theoremenv}}
\newenvironment{corollary}{\begin{corollaryenv}\rm}{\end{corollaryenv}}
\newenvironment{example}{\begin{exampleenv}\rm}{\end{exampleenv}}
\newenvironment{proposition}{\begin{propositionenv}\rm}{\end{propositionenv}}
\newenvironment{conjecture}{\begin{conjectureenv}\rm}{\end{conjectureenv}}
\newenvironment{app-lemma}{\begin{app-lemmaenv}\rm}{\end{app-lemmaenv}}

\newcommand{\bd}{\begin{definition}}
	\newcommand{\ed}{\end{definition}}
\newcommand{\bl}{\begin{lemma}}
	\newcommand{\el}{\end{lemma}}
\newcommand{\elp}{\hspace*{\fill} $\Box$
\end{lemma}}
\newcommand{\bt}{\begin{theorem}}
\newcommand{\et}{\end{theorem}}
\newcommand{\etp}{\hspace*{\fill} $\Box$
\end{theorem}}
\newcommand{\bc}{\begin{corollary}}
\newcommand{\ec}{\end{corollary}}
\newcommand{\ecp}{\hspace*{\fill} $\Box$
\end{corollary}}
\newcommand{\bcj}{\begin{conjecture}}
\newcommand{\ecj}{\end{conjecture}}

\newcommand{\be}{\begin{example}}
\newcommand{\ee}{\end{example}}
\newcommand{\eep}{\hspace*{\fill} $\Box$
\end{example}}
\newcommand{\bp}{\begin{proposition}}
\newcommand{\ep}{\end{proposition}}
\newcommand{\epp}{
\end{proposition}}

\newcommand{\wt}{\mathrm{wt}}

\ifnum\final=0
\newcommand{\mynote}[2]{{\color{#1} \marginpar{\tiny #2}}}
\newcommand{\mybignote}[2]{{\color{#1} $\langle \langle$ #2$\rangle \rangle$}}
\newcommandx{\rednote}[2][1=]{\todo[linecolor=red,backgroundcolor=red!25,bordercolor=red,#1]{#2}}
\newcommandx{\bluenote}[2][1=]{\todo[linecolor=blue,backgroundcolor=blue!25,bordercolor=blue,#1]{#2}}
\newcommandx{\yellownote}[2][1=]{\todo[linecolor=yellow,backgroundcolor=yellow!25,bordercolor=yellow,#1]{#2}}
\newcommandx{\greennote}[2][1=]{\todo[inline,linecolor=olive,backgroundcolor=green!25,bordercolor=olive,#1]{#2}}

\else
\newcommand{\mynote}[2]{}
\newcommand{\mybignote}[2]{}
\newcommand{\rednote}[2][1=]{}
\newcommand{\bluenote}[2][1=]{}
\newcommand{\greennote}[2][1=]{}
\newcommand{\yellownote}[2][1=]{}

\fi

\usepackage{adjustbox}
\usetikzlibrary{quantikz2}
\usetikzlibrary{backgrounds,fit,decorations.pathreplacing}  
\usetikzlibrary{automata} 
\usetikzlibrary{circuits.logic.CDH}
\usetikzlibrary{circuits.logic.US}
\usetikzlibrary{mindmap}
\usetikzlibrary{decorations}
\usetikzlibrary{decorations.pathmorphing}
\usetikzlibrary{arrows,decorations.pathreplacing}

\tikzset{meter/.append style={draw, inner sep=10, rectangle, font=\vphantom{A}, minimum width=30, line width=.4, path picture={\draw[black] ([shift={(.1,.3)}]path picture bounding box.south west) to[bend left=50] ([shift={(-.1,.3)}]path picture bounding box.south east);\draw[black,-latex] ([shift={(0,.1)}]path picture bounding box.south) -- ([shift={(.3,-.1)}]path picture bounding box.north);}}}

\begin{document}

\title{
Quantum dual extended Hamming code immune to collective coherent errors
}

\author{En-Jui Chang}
\email{phyenjui@gmail.com}
\affiliation{Taichung 421786, Taiwan}

\date{\today}

\begin{abstract}
Collective coherent (CC) errors are inevitable, as every physical qubit undergoes free evolution under its kinetic Hamiltonian. These errors can be more damaging than stochastic Pauli errors because they affect all qubits coherently, resulting in high-weight errors that standard quantum error-correcting (QEC) codes struggle to correct. In quantum memories and communication systems, especially when storage durations are long, CC errors often dominate over stochastic noise. Trapped-ion platforms, for example, exhibit strong CC errors with minimal stochastic Pauli components. In this work, we address the regime where immunity to CC errors, high code rate (due to limited qubit availability), and moderate distance (sufficient for correcting low-weight errors) are all essential. We construct a family of constant-excitation (CE) stabilizer codes with parameters $[[2^{r+1}, 2^r - (r+1), 3]]$. The smallest instance, the $[[8,1,3]]$ code, improves the code rate and error threshold of the best previously known CE code by factors of approximately two and four, respectively.
\end{abstract}

\maketitle

\section{Introduction}

The intrinsic Hamiltonian governing each qubit's free evolution drives coherent time dynamics, even when the qubits are idle or operating within well-controlled quantum platforms. As a result, collective coherent (CC) errors are unavoidable. Unlike stochastic Pauli errors, CC errors apply coherent, high-weight operations simultaneously across many qubits, potentially causing more severe logical errors. Although these errors do not reduce the state's purity, their structured nature makes them particularly damaging to quantum information. To solve this issue, significant progress has been made in correcting these CC errors~\cite{Chamberland2017, Bravyi2018, Debroy2018, HLRC22, Mrton2023, PhysRevA.90.062317, PhysRevA.93.042340, PhysRevA.111.052602,hwfz-c6vy}.

Although quantum memory, communication, and computation differ fundamentally from their classical counterparts, they are likely to share some common practical considerations. First, multiple sources of qubits will exist, offering varying levels of quality and cost. Second, quantum data will need to be stored for future use, often for uncertain durations. Third, quantum information may be transmitted to mobile or spatially distributed receivers whose precise locations are not always known. Fourth, transferring quantum data from memory to processing units incurs time overhead. We will address these considerations in detail later in this work.

Trapped-ion qubits~\cite{PhysRevLett.74.4091,PhysRevLett.75.4714,Hogle2023,PhysRevX.13.041052,PhysRevX.14.041017,Main2025} share architectural similarities with classical Von Neumann systems, where memory and computation are physically separated. In such platforms, qubit transport, whether through optical tweezers or ion shuttling, naturally introduces CC errors. While trapped-ion systems are known for their high-fidelity operations, they typically suffer from slower gate speeds and a limited maximum qubit count due to their inherently one-dimensional layout. As a result, the primary bottleneck lies in the number of high-quality qubits that can be reliably operated within a single device.

In addition to CC errors, quantum systems are also subject to depolarizing noise, typically modeled as stochastic Pauli errors. Two well-known quantum error-correcting (QEC) codes that address such errors are the Shor code~\cite{Shor95} and the Steane code~\cite{Ste96,PhysRevA.54.4741}. Together, they form the class of Calderbank-Shor-Steane (CSS) stabilizer codes. The Shor code offers a systematic way to increase the code distance, a measure of error-correction strength, at the expense of code rate, defined as the ratio of logical qubits to physical qubits. In contrast, the Steane code demonstrates how to improve the code rate while fixing the code distance. These two design philosophies highlight a broader principle: effective QEC must adapt to varying hardware constraints, including the availability and quality of physical qubits.

To address both CC and stochastic Pauli errors, QEC codes with constant excitation (CE) codewords are essential for reliable quantum memory and computation.

In our previous work~\cite{PhysRevA.111.052602}, we made significant progress in increasing the code distance without sacrificing too much code rate. The family of $[[2(w+1)(w+K), K]]$ CE codes, referred to as \textit{w}-codes, can be interpreted as $[[2(w+1)(w+K), K, d']]$ Shor-like codes with distance $d'=w+1$ for $K=1$, $d'=2$ for $K\ge 2$. Here, the parameter \textit{w} indicates the ability to correct weight-\textit{w} amplitude-damping (AD) errors, though not weight-\textit{w} Pauli errors. For conventional $[[n, k, d]]$ stabilizer codes designed to correct Pauli errors, up to $\lfloor\frac{d}{2}\rfloor$ errors can be detected and $\lfloor\frac{d-1}{2}\rfloor$ errors can be corrected.

In contrast, the present work prioritizes code rate over distance, targeting scenarios where only a limited number of high-quality qubits are available and where correcting weight-1 Pauli errors is sufficient. These qubits may serve as quantum memory or as carriers in quantum communication, where timing information is often inaccessible, making it impractical to correct CC errors simply by reversing coherent evolution. As a result, we aim to design Steane-like CE codes optimized for memory and communication applications, accepting reduced flexibility in code distance in exchange for higher encoding efficiency.

Before constructing our CE codes, we first examine how existing CE codes are designed. The simplest example is the $[[2,1,0]]$ complemented dual-rail (CDR) code~\cite{Knill2001}, which provides no error protection. A major class of CE codes, including our previous work~\cite{PhysRevA.111.052602}, is based on concatenating a standard outer stabilizer code with an inner CDR code. This approach underlies several known constructions, such as the $[[8,1,2]]$~\cite{Ouyang2021}, $[[18,1,3]]$, and $[[32,1,4]]$ CE codes, each requiring twice the number of qubits compared to their non-CE stabilizer counterparts.

Movassagh and Ouyang~\cite{Movassagh2024} proposed the $[[15,1,4]]$ CE code by extending beyond the stabilizer framework. While it satisfies the QEC conditions, it lacks a clear method for syndrome extraction and conditional recovery operations. Another example, the $[[5,1,2]]$ CE code described as Example 6 in~\cite{HLRC22}, appears distinct but is effectively a $[[4,1,2]]$ CE code that still relies on concatenation with the inner CDR code.

From these observations, especially from~\cite{HLRC22}, we recognize that the inner CDR code effectively acts as a two-qubit repetition code, which increases the overall code distance relative to the outer code. This suggests that many prior constructions rely on overly strong outer codes, potentially limiting code rate unnecessarily.

As a result, we propose designing an outer code that does not differentiate between $X$ and $Z$ errors, relying instead on the inner CDR code to enable effective discrimination between these error types. In this work, we introduce a new family of CE stabilizer codes with parameters $[[2^{r+1}, 2^{r} - (r+1), 3]]$. These codes are constructed by encoding classical extended Hamming codes $[2^r, 2^r - (r+1), 4]$~\cite{Hamming1950} into CE subspaces using an inner CDR code.

The smallest instance in this family is the $[[8,1,3]]$ code, which uses fewer qubits than previously known CE codes, whether stabilizer or non-stabilizer. Moreover, this new construction improves the asymptotic code rate to $\frac{1}{2}$. A summary comparison is provided in Table~\ref{table:CE_QECCs}.

\begin{table}[ht]
\caption{Comparison of representative CE QEC codes. Each $[[n,k,d]]$ code encodes $k$ logical qubits into $n$ physical qubits with distance $d$, allowing correction of up to $\lfloor\frac{d-1}{2}\rfloor$ and detection of up to $\lfloor\frac{d}{2}\rfloor$ Pauli errors. The rate is defined as the ratio $\frac{k}{n}$, indicating the encoding efficiency. The symbol $d'=d$ for $K=1$, otherwise $d'=2$ for $K\ge 2$.}
\label{table:CE_QECCs}
\begin{tabularx}{\linewidth}{ c  c  c  c  c }
 \hline
 \hline
 CE QEC codes & \hspace{0.4cm} & Rate & \hspace{0.4cm} & Advantages \\[1ex]
 \hline
 $[[5,1,2]]$~\cite{HLRC22} & &$\frac{1}{5}$ & & \textrm{CSS code} \\[2ex]
 $[[15,1,4]]$~\cite{Movassagh2024} & &$\frac{1}{15}$ & & \textrm{Supports qudits} \\[2ex]
 $[[2d(K+d-1),K, d']]$~\cite{PhysRevA.111.052602} & &$\frac{1}{2d}\frac{K}{K+d-1}$ & & \textrm{CSS code} \\[1ex]
 (Previous work) & & $\in [\frac{1}{2d^2},\frac{1}{2d}]$ & & \\[2ex]
 $[[2^{r+1},2^{r}-(r+1), 3]]$ & & $\frac{1}{2}-\frac{r+1}{2^{r+1}}$ & & \textrm{Highest rate} \\[1ex]
 (This work) & & $\in [\frac{1}{8},\frac{1}{2}]$ & & \\
 \hline
 \hline
\end{tabularx}
\end{table}

As expected, the dual extended Hamming codes presented in this work are inherently immune to CC errors while offering a high code rate. The worst-case code rate in this family is $\frac{1}{8}$, and the asymptotic rate approaches $\frac{1}{2}$. Notably, this worst-case rate already matches the best asymptotic performance achieved in our previous work~\cite{PhysRevA.111.052602}. These codes are particularly well-suited for scenarios where only a limited number of high-quality qubits are available on a single chip.

\section{Preliminary}
Pauli operators for a qubit have eigenvalues $\pm 1$. Consequently, measuring a Pauli operator yields binary outcomes $\{0,1\}$, corresponding to the eigenvalues $+1$ and $-1$, respectively. The three Pauli matrices are defined as $X=\left(\begin{array}{cc}
0 & 1\\
1 & 0
\end{array}\right)$, $Y=\left(\begin{array}{cc}
0 & -\mathrm{i}\\
\mathrm{i} & 0
\end{array}\right)$, and $Z=\left(\begin{array}{cc}
1 & 0\\
0 & {-1}
\end{array}\right)$, and the identity operator is given by $I=\left(\begin{array}{cc}
1 & 0\\
0 & 1
\end{array}\right)$.

For an $N$-qubit system, an $N$-fold Pauli operator is the tensor product of $N$ individual Pauli operators. By convention, the tensor product symbol $\otimes$ is often omitted when the context is clear. For example, one may write
\begin{equation*}
    X\otimes Y\otimes Z\otimes I= XYZI = X_{0} Y_{1} Z_{2},
\end{equation*}
where the subscripts indicate the qubits on which the operators act.

Let $\mathcal{G}_N$ denote the $N$-fold Pauli group, which consists of all $N$-fold tensor products of the Pauli matrices (and the identity), together with a phase factor chosen from \(\{\pm 1, \pm \mathrm{i}\}\).

In the context of an $[[N, K, D]]$ stabilizer code, the stabilizer group \(\mathcal{S} = \langle g_0, \dots, g_{N-K-1} \rangle \subset \mathcal{G}_{N}\) s generated by $N-K$ mutually commuting elements in $\mathcal{G}_N$. Measuring these stabilizer generators allows one to extract error syndromes without disturbing the encoded codewords.

Logical (Pauli) operators are elements of the normalizer $\mathcal{N}(\mathcal{S}) \subset \mathcal{G}_N$, and are typically generated by the set $\big\langle \overline{X}_0,\dots, \overline{X}_{K-1}$, $\overline{Z}_0,\dots, \overline{Z}_{K-1} \big\rangle \subset \mathcal{N}(\mathcal{S})$.

Here, the overline is used to distinguish logical operators from their physical counterparts. For instance, a physical operator $P \in {X,Y,Z}$ (or a state $\ket{\psi}$) is encoded as the logical operator $\overline{P}$ (or the logical state $\overline{\ket{\psi}}$).

To ensure that logical operations preserve the code space, each logical operator in $\mathcal{N}(\mathcal{S})$ must commute with every element of the stabilizer group $\mathcal{S}$.

\subsection{Composite error model}
For any positive circuit time $\Delta t \in \mathbb{R}^{+}$, an $N$-qubit system evolving under the intrinsic Hamiltonian $\hat{H}_{0}=\sum_{j=0}^{N-1}\frac{\hbar}{2}\hat{Z}_{j}$ experiences coherent evolution described by the CC channel:
\begin{align}
    \mathcal{E}_{\text{CC}}:\;\rho \mapsto& e^{-\mathrm{i}\hat{H}_{0}\Delta t} \rho e^{\mathrm{i}\hat{H}_{0}\Delta t}.
\end{align}
For simplicity, we set $\frac{\hbar}{2} = 1$.

On the other hand, the depolarizing channel acting on a single qubit is defined via the Kraus operators:
\begin{align}
    (\hat{K}_{0}, \hat{K}_{1}, \hat{K}_{2}, \hat{K}_{3}) = & (\sqrt{1-p}\hat{I}, \sqrt{\frac{p}{3}}\hat{X}, \sqrt{\frac{p}{3}}\hat{Y}, \sqrt{\frac{p}{3}}\hat{Z}).
\end{align}

For an $N$-qubit system, the depolarizing channel is given by
\begin{align}
    \mathcal{E}_{\text{P}}:\; \rho \mapsto&\;  \sum_{j\in\{0,1,2,3\}^{N}}(\bigotimes_{k=0}^{N-1}\hat{K}_{j_{k}})\rho (\bigotimes_{k=0}^{N-1}\hat{K}_{j_{k}}^{\dagger}).
\end{align}

Here, the multi-index $j = (j_0, j_1, \ldots, j_{N-1})$ is used to denote the particular Kraus operator acting on each qubit, and we define the weight $\wt(j)$ as the number of non-zero elements in $j$. Thus, the operator $\bigotimes_{k=0}^{N-1} \hat{K}_{j_k}$ corresponds to a depolarizing error of weight $\wt(j)$.

There are two possible orderings for the overall error model: either $\mathcal{E}_{\text{CC}} \circ \mathcal{E}_{\text{P}}$ or $\mathcal{E}_{\text{P}} \circ \mathcal{E}_{\text{CC}}$. In this work, we focus on the former to illustrate the main idea.

\subsection{Classical extended Hamming code}
Before presenting the complete QEC code, we first examine the outer classical code. For a block of $2^r$ (qu)bits, at most $r+1$ parity checks are sufficient to uniquely identify a single phase error. To illustrate this, we express the classical code using typical stabilizers from a quantum code designed to correct a $Z$ error:
\begin{subequations}
\begin{align}
    g_{0}^{X} &= \bigotimes_{j=2^{r-1}}^{2^{r}-1} X_{j},\\[1mm]
    g_{i}^{X} &= \bigotimes_{j=0}^{2^{i}-1}\bigotimes_{k=0}^{2^{r-i}-1} X_{j(2^{r-i})+k},\quad i\in\{1,\dots,r\}.
\end{align}
\end{subequations}
Similarly, the stabilizers for correcting an $X$ error are given by
\begin{subequations}
\begin{align}
    g_{0}^{Z} &= \bigotimes_{j=2^{r-1}}^{2^{r}-1} Z_{2^{r}-1-j},\\[1mm]
    g_{i}^{Z} &= \bigotimes_{j=0}^{2^{i}-1}\bigotimes_{k=0}^{2^{r-i}-1} Z_{2^{r}-1-(j(2^{r-i})+k)},\quad i\in\{1,\dots,r\}.
\end{align}
\end{subequations}

The new set of $r+1$ stabilizers that combine these two types is
\begin{align}
    g_{i} = g_{i}^{X}\otimes g_{i}^{Z},\quad i\in\{0,\dots,r\},
\end{align}
which can detect a single error of one type. In fact, this code is of distance $2$, e.g. the $X_{2^{r-1}-1}Z_{2^{r-1}}$ operator cannot be detected.

Next, we incorporate the additional CDR code to distinguish between $Z$ and $X$ errors.

\section{Quantum dual extended Hamming code}
To construct a useful QEC code, we introduce an inner CDR code with stabilizer $-ZZ$ and logical operators defined by
\[
\overline{X} = XX, \quad \overline{Z} = ZI.
\]
The concatenated stabilizers are then given by
\begin{subequations}
\begin{align}
    g_{0} =& \bigotimes_{j=2^{r-1}}^{2^{r}-1} \; X_{j}\, Z_{2^{r}-1-j}\, X_{j+2^{r}},\\[1mm]
    g_{i} =& \bigotimes_{j=0}^{2^{i}-1} \bigotimes_{k=0}^{2^{r-i}-1} \; X_{j(2^{r-i})+k}\, Z_{2^{r}-1-(j(2^{r-i})+k)}\, X_{j(2^{r-i})+k+2^{r}},\notag\\
    & i\in\{1,\dots,r\},\\
    \text{and}\notag\\
    g_{i} =& -Z_{i-(r+1)}\, Z_{i-(r+1)+2^{r}},\quad i\in\{r+1,\dots,r+2^{r}\}.
\end{align}
\end{subequations}
Since there are $N = 2^{r+1}$ physical qubits and a total of $|\mathcal{S}| = 2^{r} + r + 1$ stabilizers, the code encodes
\[
K = 2^{r} - (r+1)
\]
logical qubits. The aforementioned weight-2 operator $X_{2^{r-1}-1}Z_{2^{r-1}}$ is now becomes $X_{2^{r-1}-1}Z_{2^{r-1}}X_{2^{r-1}-1+2^r}$, and therefore this code is of distance $3$ rather than the desired distance $4$ of the classical extended Hamming code.

\subsection*{Smallest example}
The smallest instance of this construction is the $[[8,1,3]]$ CE code corresponding to $r = 3$, the encoding circuit is shown in Fig.~\ref{fig:enc[[8,1,3]]}. Its stabilizers are:
\begin{subequations}
\begin{align}
    g_{0} &= ZZXXIIXX,\\
    g_{1} &= XXZZXXII,\\
    g_{2} &= XZXZXIXI,\\
    g_{3} &= -ZIIIZIII,\\
    g_{4} &= -IZIIIZII,\\
    g_{5} &= -IIZIIIZI,\\
    g_{6} &= -IIIZIIIZ,
\end{align}
\end{subequations}
and the logical Pauli operators are given by
\begin{subequations}
\begin{align}
    \overline{X} &= IZZYIIIX,\\[1mm]
    \overline{Z} &= ZZZZIIII.
\end{align}
\end{subequations}

\begin{figure}[ht]
    \begin{tikzcd}[row sep=0.1cm, column sep=0.3cm]
    \ket{+}&\ctrl{3}&\qw&\ctrl{2}&\gate{\hat{S}\hat{H}\hat{T}^{\dagger}}&\ctrl{4}&\qw&\qw&\qw&\rstick[8]{$\overline{\ket{\psi}}$}\\
    \hat{S}\hat{X}\ket{\psi}&\qw&\ctrl{1}&\targ{}&\gate{\hat{S}\hat{H}\hat{T}^{\dagger}}&\qw&\ctrl{4}&\qw&\qw&\\
    \ket{0}&\qw&\targ{}&\targ{}&\gate{\hat{S}\hat{H}\hat{T}^{\dagger}}&\qw&\qw&\ctrl{4}&\qw&\\
    \ket{0}&\targ{}&\qw&\qw&\gate{\hat{S}\hat{H}\hat{T}^{\dagger}}&\qw&\qw&\qw&\ctrl{4}&\\
    \ket{1}&\qw&\qw&\qw&\qw&\targ{}&\qw&\qw&\qw&\\
    \ket{1}&\qw&\qw&\qw&\qw&\qw&\targ{}&\qw&\qw&\\
    \ket{1}&\qw&\qw&\qw&\qw&\qw&\qw&\targ{}&\qw&\\
    \ket{1}&\qw&\qw&\qw&\qw&\qw&\qw&\qw&\targ{}&
    \end{tikzcd}
    \caption{Encoding circuit for the $[[8,1,3]]$ CE QEC code. This stabilizer code encodes one logical qubit into eight physical qubits, offering protection against all weight-1 Pauli errors. It achieves a high encoding rate and is designed to be immune to CC errors.}\label{fig:enc[[8,1,3]]}
\end{figure}
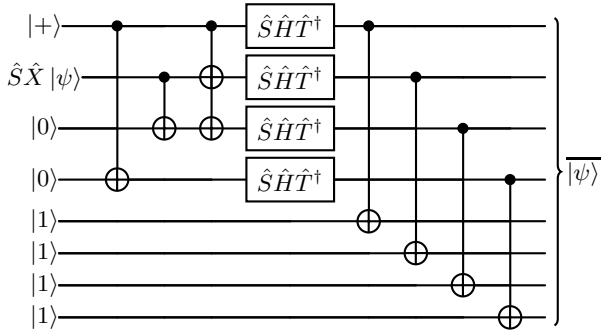

Suppose the input state is $\ket{\psi}=\alpha\ket{0}+\beta\ket{1}$. We first encode it into the logical four-qubit outer code $\ket{\psi}_{\textrm{outer}}^{(r=2)}$,  defined as follows:
\begin{align*}
    \ket{0}_{\mathrm{outer}}^{(r=2)}\coloneqq&\frac{1}{\sqrt{2}}\left(\ket{+Y}\ket{-Y}^{\otimes 2}\ket{+Y}+\ket{-Y}\ket{+Y}^{\otimes 2}\ket{-Y}\right),\\
    \ket{1}_{\mathrm{outer}}^{(r=2)}\coloneqq&\frac{1}{\sqrt{2}}\left(\ket{+Y}^{\otimes 4}-\ket{-Y}^{\otimes 4}\right),
\end{align*}
and then concatenate this outer code with the inner CDR code,
\begin{align}
    &\overline{\ket{\psi}}^{(r=2)}\notag\\
    \coloneqq&\bigotimes_{i=0}^{3}(\ket{0}_{i}\bra{0}\otimes\hat{I}_{i+4}+\ket{1}_{i}\bra{1}\otimes\hat{X}_{i+4})\ket{\psi}_{\mathrm{outer}}^{(r=2)}\ket{1}^{\otimes 4}.
\end{align}

\begin{table}[ht]
\caption{Syndrome extraction table for the CE $[[8,1,3]]$ code. Note that some $Z$ errors share the same syndrome, indicating degeneracy.}
\label{tab:syndrome_813}
\begin{tabularx}{\linewidth}{ c  c  c  c  c  c  c  c  c }
    \hline
    \hline
    Error & Syndrome & \hspace{0.4cm} & Error & Syndrome & \hspace{0.4cm} & Error & Syndrome\\
    \hline
    $X_{0}$  & 1001000 & & $Y_{0}$  & 1111000 & & $Z_{0}$  & 0110000\\
    $X_{1}$  & 1010100 & & $Y_{1}$  & 1110100 & & $Z_{1}$  & 0100000\\
    $X_{2}$  & 0100010 & & $Y_{2}$  & 1110010 & & $Z_{2}$  & 1010000\\
    $X_{3}$  & 0110001 & & $Y_{3}$  & 1110001 & & $Z_{3}$  & 1000000\\
    $X_{4}$  & 0001000 & & $Y_{4}$  & 0111000 & & $Z_{4}$  & 0110000\\
    $X_{5}$  & 0000100 & & $Y_{5}$  & 0100100 & & $Z_{5}$  & 0100000\\
    $X_{6}$  & 0000010 & & $Y_{6}$  & 1010010 & & $Z_{6}$  & 1010000\\
    $X_{7}$  & 0000001 & & $Y_{7}$  & 1000001 & & $Z_{7}$  & 1000000\\
    \hline
    \hline
\end{tabularx}
\end{table}

By measuring the error syndromes and correcting the corresponding weight-1 Pauli errors using the lookup table in Table~\ref{tab:syndrome_813}, the worst-case fidelity is given by the total probability of all correctable error events:
\begin{align}
    \mathcal{F}(p)=&(1-p)^{8}+\binom{8}{1}p(1-p)^{7}+\binom{8}{2}p^{2}(1-p)^{6}+O(p^{3}),\notag\\
    =&1-48 p^{2}+O(p^{3}).
\end{align}
We can estimate the code capacity threshold $p_{\text{th}}$ by requiring the fidelity to remain above $1-p$, yielding:
\begin{align}
    1-48p_{\text{th}}^{2}\ge&1-p_{\text{th}},\notag\\
    \Rightarrow \quad p_{\text{th}}\le&\frac{1}{48}\approx 2.083\times 10^{-2}.
\end{align}
More generally, for an $n$-qubit QEC code that corrects weight-1 Pauli errors in a similar manner, the threshold can be estimated from:
\begin{align}
    \sum_{i=0}^{2} \binom{n}{i} p_{\text{th}}^i (1 - p_{\text{th}})^{n - i} + O(p_{\text{th}}^3) &\ge 1 - p_{\text{th}},\notag\\
    \Rightarrow \quad p_{\text{th}} &\le \frac{1}{n(n - 2)}.
\end{align}
This result shows that the error threshold scales approximately as the inverse square of the number of physical qubits. Therefore, improving the code rate also leads to an enhancement in the code capacity threshold, assuming ideal gates and syndrome extraction. Compared to the $[[15,1,4]]$ non-stabilizer CE code from~\cite{Movassagh2024}, the improvement in the estimated threshold is approximately:
\begin{align*}
    \frac{15\times 13}{8\times 6}\approx 4.063.
\end{align*}
We do not consider gate errors in this threshold comparison, as the $[[15,1,4]]$ non-stabilizer CE code lacks a defined quantum circuit for syndrome extraction and recovery. Including gate errors in our analysis while the compared code has no corresponding circuit implementation would result in an unfair comparison.

\section{Conclusion}
In this work, we considered scenarios where only a limited number of high-quality qubits are available, such as in quantum memories and communication systems subject to unavoidable CC errors and minor stochastic Pauli errors. To address these challenges, we introduced a new family of CE stabilizer codes with parameters $[[2^{r+1}, 2^r - (r+1), 3]]$, designed to mitigate both error types while optimizing code rate. The smallest member of this family, the $[[8,1,3]]$ CE code, reduces the required qubit count by half compared to the previously known CE code and improves the error threshold by a factor of four.

The primary trade-offs introduced to achieve higher code rates in this work are as follows:
\begin{itemize}
    \item We adopt a Steane-like construction instead of the Shor-based approach used in our previous work, which sacrifices the flexibility to easily increase the code distance. However, this limitation is mitigated by our previously developed high-distance CE codes.
    \item The use of higher-weight stabilizers may increase the time required for syndrome measurements, potentially leading to a higher accumulation of CC errors. Fortunately, the CE codewords are inherently immune to such additional CC errors, preserving their robustness in this setting.
\end{itemize}

These results demonstrate the potential of our construction as a foundation for more resource-efficient quantum memories and communication systems.


\begin{thebibliography}{22}
\providecommand{\natexlab}[1]{#1}
\providecommand{\url}[1]{\texttt{#1}}
\expandafter\ifx\csname urlstyle\endcsname\relax
  \providecommand{\doi}[1]{doi: #1}\else
  \providecommand{\doi}{doi: \begingroup \urlstyle{rm}\Url}\fi

\bibitem[Chamberland et~al.(2017)Chamberland, Wallman, Beale, and Laflamme]{Chamberland2017}
Christopher Chamberland, Joel Wallman, Stefanie Beale, and Raymond Laflamme.
\newblock Hard decoding algorithm for optimizing thresholds under general markovian noise.
\newblock \emph{Physical Review A}, 95\penalty0 (4), April 2017.
\newblock ISSN 2469-9934.
\newblock \doi{10.1103/physreva.95.042332}.
\newblock URL \url{http://dx.doi.org/10.1103/PhysRevA.95.042332}.

\bibitem[Bravyi et~al.(2018)Bravyi, Englbrecht, K\"{o}nig, and Peard]{Bravyi2018}
Sergey Bravyi, Matthias Englbrecht, Robert K\"{o}nig, and Nolan Peard.
\newblock Correcting coherent errors with surface codes.
\newblock \emph{npj Quant. Inf.}, 4\penalty0 (1), October 2018.
\newblock ISSN 2056-6387.
\newblock \doi{10.1038/s41534-018-0106-y}.
\newblock URL \url{http://dx.doi.org/10.1038/s41534-018-0106-y}.

\bibitem[Debroy et~al.(2018)Debroy, Li, Newman, and Brown]{Debroy2018}
Dripto~M. Debroy, Muyuan Li, Michael Newman, and Kenneth~R. Brown.
\newblock Stabilizer slicing: Coherent error cancellations in low-density parity-check stabilizer codes.
\newblock \emph{Physical Review Letters}, 121\penalty0 (25), December 2018.
\newblock ISSN 1079-7114.
\newblock \doi{10.1103/physrevlett.121.250502}.
\newblock URL \url{http://dx.doi.org/10.1103/PhysRevLett.121.250502}.

\bibitem[Hu et~al.(2022)Hu, Liang, Rengaswamy, and Calderbank]{HLRC22}
Jingzhen Hu, Qingzhong Liang, Narayanan Rengaswamy, and Robert Calderbank.
\newblock Mitigating coherent noise by balancing weight-2 z-stabilizers.
\newblock \emph{IEEE Transactions on Information Theory}, 68\penalty0 (3):\penalty0 1795--1808, 2022.
\newblock \doi{10.1109/TIT.2021.3130155}.

\bibitem[M\'{a}rton and Asb\'{o}th(2023)]{Mrton2023}
\'{A}ron M\'{a}rton and J\'{a}nos~K. Asb\'{o}th.
\newblock Coherent errors and readout errors in the surface code.
\newblock \emph{Quantum}, 7:\penalty0 1116, September 2023.
\newblock ISSN 2521-327X.
\newblock \doi{10.22331/q-2023-09-21-1116}.
\newblock URL \url{http://dx.doi.org/10.22331/q-2023-09-21-1116}.

\bibitem[Ouyang(2014)]{PhysRevA.90.062317}
Yingkai Ouyang.
\newblock Permutation-invariant quantum codes.
\newblock \emph{Phys. Rev. A}, 90:\penalty0 062317, Dec 2014.
\newblock \doi{10.1103/PhysRevA.90.062317}.
\newblock URL \url{https://link.aps.org/doi/10.1103/PhysRevA.90.062317}.

\bibitem[Ouyang and Fitzsimons(2016)]{PhysRevA.93.042340}
Yingkai Ouyang and Joseph Fitzsimons.
\newblock Permutation-invariant codes encoding more than one qubit.
\newblock \emph{Phys. Rev. A}, 93:\penalty0 042340, Apr 2016.
\newblock \doi{10.1103/PhysRevA.93.042340}.
\newblock URL \url{https://link.aps.org/doi/10.1103/PhysRevA.93.042340}.

\bibitem[Chang and Lai(2025)]{PhysRevA.111.052602}
En-Jui Chang and Ching-Yi Lai.
\newblock High-rate amplitude-damping shor codes with immunity to collective coherent errors.
\newblock \emph{Phys. Rev. A}, 111:\penalty0 052602, May 2025.
\newblock \doi{10.1103/PhysRevA.111.052602}.
\newblock URL \url{https://link.aps.org/doi/10.1103/PhysRevA.111.052602}.

\bibitem[Chang(2025)]{hwfz-c6vy}
En-Jui Chang.
\newblock High-rate extended binomial codes for multiqubit encoding.
\newblock \emph{Phys. Rev. A}, 112:\penalty0 032419, Sep 2025.
\newblock \doi{10.1103/hwfz-c6vy}.
\newblock URL \url{https://link.aps.org/doi/10.1103/hwfz-c6vy}.

\bibitem[Cirac and Zoller(1995)]{PhysRevLett.74.4091}
J.~I. Cirac and P.~Zoller.
\newblock Quantum computations with cold trapped ions.
\newblock \emph{Phys. Rev. Lett.}, 74:\penalty0 4091--4094, May 1995.
\newblock \doi{10.1103/PhysRevLett.74.4091}.
\newblock URL \url{https://link.aps.org/doi/10.1103/PhysRevLett.74.4091}.

\bibitem[Monroe et~al.(1995)Monroe, Meekhof, King, Itano, and Wineland]{PhysRevLett.75.4714}
C.~Monroe, D.~M. Meekhof, B.~E. King, W.~M. Itano, and D.~J. Wineland.
\newblock Demonstration of a fundamental quantum logic gate.
\newblock \emph{Phys. Rev. Lett.}, 75:\penalty0 4714--4717, Dec 1995.
\newblock \doi{10.1103/PhysRevLett.75.4714}.
\newblock URL \url{https://link.aps.org/doi/10.1103/PhysRevLett.75.4714}.

\bibitem[Hogle et~al.(2023)Hogle, Dominguez, Dong, Leenheer, McGuinness, Ruzic, Eichenfield, and Stick]{Hogle2023}
C.~W. Hogle, D.~Dominguez, M.~Dong, A.~Leenheer, H.~J. McGuinness, B.~P. Ruzic, M.~Eichenfield, and D.~Stick.
\newblock High-fidelity trapped-ion qubit operations with scalable photonic modulators.
\newblock \emph{npj Quantum Information}, 9\penalty0 (1), July 2023.
\newblock ISSN 2056-6387.
\newblock \doi{10.1038/s41534-023-00737-1}.
\newblock URL \url{http://dx.doi.org/10.1038/s41534-023-00737-1}.

\bibitem[Moses et~al.(2023)Moses, Baldwin, Allman, Ancona, Ascarrunz, Barnes, Bartolotta, Bjork, Blanchard, Bohn, Bohnet, Brown, Burdick, Burton, Campbell, Campora, Carron, Chambers, Chan, Chen, Chernoguzov, Chertkov, Colina, Curtis, Daniel, DeCross, Deen, Delaney, Dreiling, Ertsgaard, Esposito, Estey, Fabrikant, Figgatt, Foltz, Foss-Feig, Francois, Gaebler, Gatterman, Gilbreth, Giles, Glynn, Hall, Hankin, Hansen, Hayes, Higashi, Hoffman, Horning, Hout, Jacobs, Johansen, Jones, Karcz, Klein, Lauria, Lee, Liefer, Lu, Lucchetti, Lytle, Malm, Matheny, Mathewson, Mayer, Miller, Mills, Neyenhuis, Nugent, Olson, Parks, Price, Price, Pugh, Ransford, Reed, Roman, Rowe, Ryan-Anderson, Sanders, Sedlacek, Shevchuk, Siegfried, Skripka, Spaun, Sprenkle, Stutz, Swallows, Tobey, Tran, Tran, Vogt, Volin, Walker, Zolot, and Pino]{PhysRevX.13.041052}
S.~A. Moses, C.~H. Baldwin, M.~S. Allman, R.~Ancona, L.~Ascarrunz, C.~Barnes, J.~Bartolotta, B.~Bjork, P.~Blanchard, M.~Bohn, J.~G. Bohnet, N.~C. Brown, N.~Q. Burdick, W.~C. Burton, S.~L. Campbell, J.~P. Campora, C.~Carron, J.~Chambers, J.~W. Chan, Y.~H. Chen, A.~Chernoguzov, E.~Chertkov, J.~Colina, J.~P. Curtis, R.~Daniel, M.~DeCross, D.~Deen, C.~Delaney, J.~M. Dreiling, C.~T. Ertsgaard, J.~Esposito, B.~Estey, M.~Fabrikant, C.~Figgatt, C.~Foltz, M.~Foss-Feig, D.~Francois, J.~P. Gaebler, T.~M. Gatterman, C.~N. Gilbreth, J.~Giles, E.~Glynn, A.~Hall, A.~M. Hankin, A.~Hansen, D.~Hayes, B.~Higashi, I.~M. Hoffman, B.~Horning, J.~J. Hout, R.~Jacobs, J.~Johansen, L.~Jones, J.~Karcz, T.~Klein, P.~Lauria, P.~Lee, D.~Liefer, S.~T. Lu, D.~Lucchetti, C.~Lytle, A.~Malm, M.~Matheny, B.~Mathewson, K.~Mayer, D.~B. Miller, M.~Mills, B.~Neyenhuis, L.~Nugent, S.~Olson, J.~Parks, G.~N. Price, Z.~Price, M.~Pugh, A.~Ransford, A.~P. Reed, C.~Roman, M.~Rowe, C.~Ryan-Anderson, S.~Sanders, J.~Sedlacek, P.~Shevchuk, P.~Siegfried,
  T.~Skripka, B.~Spaun, R.~T. Sprenkle, R.~P. Stutz, M.~Swallows, R.~I. Tobey, A.~Tran, T.~Tran, E.~Vogt, C.~Volin, J.~Walker, A.~M. Zolot, and J.~M. Pino.
\newblock A race-track trapped-ion quantum processor.
\newblock \emph{Phys. Rev. X}, 13:\penalty0 041052, Dec 2023.
\newblock \doi{10.1103/PhysRevX.13.041052}.
\newblock URL \url{https://link.aps.org/doi/10.1103/PhysRevX.13.041052}.

\bibitem[Schwerdt et~al.(2024)Schwerdt, Peleg, Shapira, Priel, Florshaim, Gross, Zalic, Afek, Akerman, Stern, Kish, and Ozeri]{PhysRevX.14.041017}
David Schwerdt, Lee Peleg, Yotam Shapira, Nadav Priel, Yanay Florshaim, Avram Gross, Ayelet Zalic, Gadi Afek, Nitzan Akerman, Ady Stern, Amit~Ben Kish, and Roee Ozeri.
\newblock Scalable architecture for trapped-ion quantum computing using rf traps and dynamic optical potentials.
\newblock \emph{Phys. Rev. X}, 14:\penalty0 041017, Oct 2024.
\newblock \doi{10.1103/PhysRevX.14.041017}.
\newblock URL \url{https://link.aps.org/doi/10.1103/PhysRevX.14.041017}.

\bibitem[Main et~al.(2025)Main, Drmota, Nadlinger, Ainley, Agrawal, Nichol, Srinivas, Araneda, and Lucas]{Main2025}
D.~Main, P.~Drmota, D.~P. Nadlinger, E.~M. Ainley, A.~Agrawal, B.~C. Nichol, R.~Srinivas, G.~Araneda, and D.~M. Lucas.
\newblock Distributed quantum computing across an optical network link.
\newblock \emph{Nature}, February 2025.
\newblock ISSN 1476-4687.
\newblock \doi{10.1038/s41586-024-08404-x}.
\newblock URL \url{http://dx.doi.org/10.1038/s41586-024-08404-x}.

\bibitem[Calderbank and Shor(1996)]{Shor95}
A.~R. Calderbank and Peter~W. Shor.
\newblock Good quantum error-correcting codes exist.
\newblock \emph{Phys. Rev. A}, 54:\penalty0 1098--1105, Aug 1996.
\newblock \doi{10.1103/PhysRevA.54.1098}.
\newblock URL \url{https://link.aps.org/doi/10.1103/PhysRevA.54.1098}.

\bibitem[Steane(1996{\natexlab{a}})]{Ste96}
A.~M. Steane.
\newblock Multiple particle interference and quantum error correction.
\newblock \emph{Proc. R. Soc. London A}, 452:\penalty0 2551--2576, 1996{\natexlab{a}}.

\bibitem[Steane(1996{\natexlab{b}})]{PhysRevA.54.4741}
A.~M. Steane.
\newblock Simple quantum error-correcting codes.
\newblock \emph{Phys. Rev. A}, 54:\penalty0 4741--4751, Dec 1996{\natexlab{b}}.
\newblock \doi{10.1103/PhysRevA.54.4741}.
\newblock URL \url{https://link.aps.org/doi/10.1103/PhysRevA.54.4741}.

\bibitem[Knill et~al.(2001)Knill, Laflamme, and Milburn]{Knill2001}
E.~Knill, R.~Laflamme, and G.~J. Milburn.
\newblock A scheme for efficient quantum computation with linear optics.
\newblock \emph{Nature}, 409\penalty0 (6816):\penalty0 46–52, January 2001.
\newblock ISSN 1476-4687.
\newblock \doi{10.1038/35051009}.
\newblock URL \url{http://dx.doi.org/10.1038/35051009}.

\bibitem[Ouyang(2021)]{Ouyang2021}
Yingkai Ouyang.
\newblock Avoiding coherent errors with rotated concatenated stabilizer codes.
\newblock \emph{npj Quantum Information}, 7\penalty0 (1), June 2021.
\newblock ISSN 2056-6387.
\newblock \doi{10.1038/s41534-021-00429-8}.
\newblock URL \url{http://dx.doi.org/10.1038/s41534-021-00429-8}.

\bibitem[Movassagh and Ouyang(2024)]{Movassagh2024}
Ramis Movassagh and Yingkai Ouyang.
\newblock Constructing quantum codes from any classical code and their embedding in ground space of local hamiltonians.
\newblock \emph{Quantum}, 8:\penalty0 1541, November 2024.
\newblock ISSN 2521-327X.
\newblock \doi{10.22331/q-2024-11-27-1541}.
\newblock URL \url{http://dx.doi.org/10.22331/q-2024-11-27-1541}.

\bibitem[Hamming(1950)]{Hamming1950}
R.~W. Hamming.
\newblock Error detecting and error correcting codes.
\newblock \emph{The Bell System Technical Journal}, 29\penalty0 (2):\penalty0 147--160, 1950.
\newblock \doi{10.1002/j.1538-7305.1950.tb00463.x}.

\end{thebibliography}
\end{document}